# Water window soft x-ray source enabled by 25-W few-cycle mid-IR OPCPA at 100 kHz

Justinas Pupeikis*, Pierre-Alexis Chevreuil, Nicolas Bigler, Lukas Gallmann, Christopher R. Phillips and Ursula Keller

Department of Physics, Institute of Quantum Electronics, ETH Zurich, Switzerland
*Corresponding author: pupeikis@phys.ethz.ch

**Abstract**

Coherent soft x-ray (SXR) sources enable fundamental studies in the important water window spectral region. Until now, such sources have been limited to repetition rates of 1 kHz or less, which limits count rates and signal-to-noise ratio for a variety of experiments. SXR generation at high repetition rate has remained challenging because of the missing high-power mid-infrared (mid-IR) laser sources to drive the high-harmonic generation (HHG) process. Here we present a mid-IR optical parametric chirped pulse amplifier (OPCPA) centered at a wavelength of 2.2 μm and generating 16.5-fs pulses (2.2 oscillation cycles of the carrier wave) with 25 W of average power and a peak power exceeding 14 GW at 100-kHz pulse repetition rate. This corresponds to the highest reported peak power for high-repetition-rate mid-IR laser systems. The output of this 2.2-μm OPCPA system was used to generate a SXR continuum extending beyond 0.6 keV through HHG in a high-pressure gas cell.

**Main**

Progress in laser technology has enabled rapid development in attosecond science which led to many scientific discoveries [1,2]. Further advances in attosecond science are closely linked to high-harmonic generation (HHG) sources [3,4], and therefore to state-of-the-art laser systems to drive the HHG process into new performance frontiers. Specifically, there is currently great interest in scaling HHG sources to parameters beyond those available in conventional Ti:sapphire amplifier driven beamlines, in particular to higher photon energies and higher repetition rates. Photon energies extending up to 1.6 keV were generated at 20 Hz repetition rate [5]. Recently, multiple research groups have developed 1-kHz laser sources capable of producing coherent soft x-ray (SXR) radiation spanning up to the oxygen K-edge at 543 eV [6–8].

Such high-photon-energy sources are interesting for a variety of spectroscopic studies since core electrons can be accessed directly. For example, this enables direct probing of biological molecules in aqueous solutions [9], tracking of electronic, vibrational and rotational [10] as well as magnetization dynamics [8]. Furthermore, the high photon energies allow for the shortest probe pulses ever produced [11]. On the other hand, high repetition rates are especially important for applications limited by space-charge effects, such as the investigation of photoemission delays from surfaces [12,13].

The coherent SXR radiation in the above examples is generated via HHG. At a given intensity $I$ and carrier wavelength $\lambda$, the maximum energy of the generated photons scales with $\sim I \cdot \lambda^2$ of the driving laser field [14]. Thus, to obtain a high-energy cut-off without excessive ionization of the target, which would prevent phase-matching, mid-IR driving lasers are required. Longer driving wavelengths also give rise to higher phase-matching pressures, which increases the number of potential emitters [15]. On the other hand, the single-atom yield drops rapidly with wavelength, with a scaling of around $\sim \lambda^{-5.5}$ for a fixed energy interval [16]. This can lead to a significantly reduced HHG efficiency. Therefore, high-average power laser sources with pulses providing sufficient peak power are critical. Furthermore, for experiments where few photo-ionization events per shot are required to avoid space-charge effects [17] or to enable coincidence detection [18,19], the repetition rate of the laser determines the data acquisition time. Increased signal rates are not only beneficial for the signal-to-noise ratio (SNR) of the experiments but also enable studies that are not considered to be practical with traditional sources, such as experiments on samples with a limited lifetime due to being susceptible to contamination even under ultra-high vacuum conditions.

Recently, several 10-W-class mid-IR (>2 µm) laser systems operating at high repetition rates were reported [20–24]. These systems represent an important advance in ultrafast laser technology. However, in the context of efficient attosecond pulse generation at high photon energies, pulse duration and peak power are of critical importance. Until now the peak power of 100-kHz mid-IR laser sources was limited to 6 GW [20].

In this letter, we present a mid-IR laser system delivering pulses centered at a wavelength of 2.2 µm and with 14 GW peak power at 100 kHz repetition rate. The peak power of this laser source was sufficient to generate a SXR continuum extending beyond the water window. This first demonstration of SXR generation at 100 kHz is a key-enabling step for a new generation of attosecond technology where high repetition rates and high photon energies are provided simultaneously. In the following, we first present our new OPCPA system and then show data from SXR generation experiments in various noble gases, reaching a photon energy up to 0.6 keV using helium as a target.

Figure 1 shows the conceptual layout of our mid-IR optical parametric chirped pulse amplifier (OPCPA). The pumping architecture of our OPCPA is based on a Ti:sapphire laser oscillator which seeds an Innoslab amplifier system (A400, Amphos GmbH). In comparison to our last result [20], we implemented significant new design steps for the OPCPA system to obtain higher performance and improved stability.

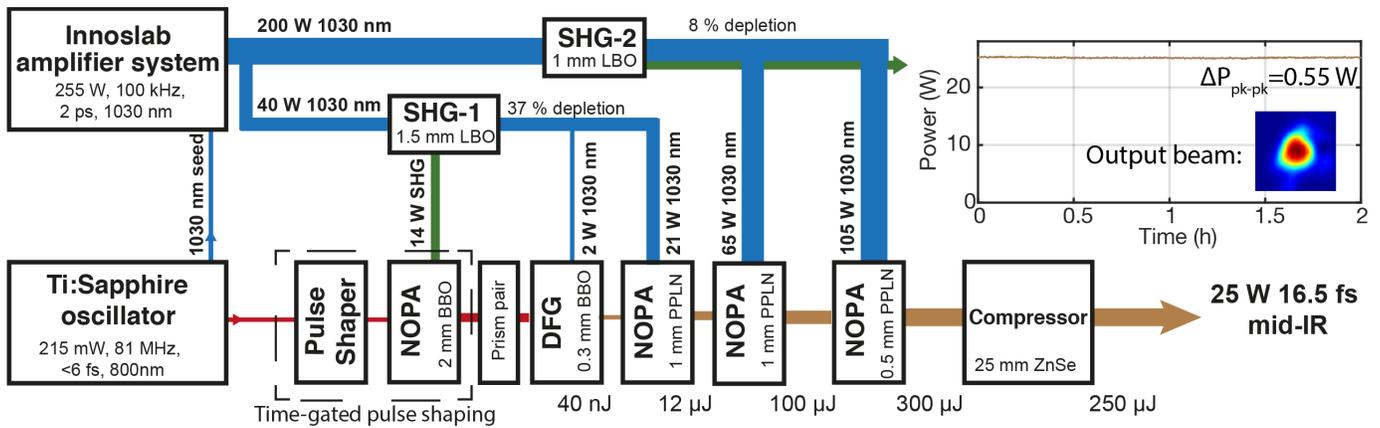

Figure 1. Optical parametric chirped-pulse amplifier (OPCPA) layout. SHG - second harmonic generation, NOPA – noncollinear optical parametric amplifier, DFG – difference frequency generation, BBO – beta barium borate, PPLN – periodically poled lithium niobate, LBO – lithium triborate, ZnSe – zinc selenide. The inset on the top right shows the long-term output stability of the system and beam profile after elliptical reshaping telescope.

First, the beam pointing of the Innoslab amplifier output is actively stabilized in using two stabilization units (TEM Messtechnik GmbH). In a first stage, the beam is stabilized on a 1-dimensional spatial Fourier filter for beam cleaning. The spatially filtered beam then enters a grating compressor where it is compressed to a duration of 2 ps. The output beam from the compressor is then spatially stabilized in a second stage to ensure a stable pump beam for the OPCPA system. The overall efficiency of this cleaning and compression unit is 61%, and the resulting total output average power is 255 W.

For the signal path of the OPCPA, we leverage a spectral-aberration-free time-gated pulse shaping technique which is applied on the low-energy seed pulses [25]. We impose a large negative group delay dispersion (GDD) of -2400 fs$^2$ at a center wavelength of 730 nm in conjunction with a negative third-order dispersion (TOD) contribution of -10400 fs$^3$. Furthermore, the Ti:sapphire spectrum is flattened via the programmable amplitude shaping capability of the pulse shaper. After shaping, the seed waveform is time-gated and amplified in a noncollinear optical parametric amplifier (NOPA) based on a 2-mm-long beta barium borate (BBO) crystal in a type-1 nonlinear interaction geometry. The 515-nm NOPA pump is generated via second harmonic generation (SHG) of a portion of the 1030-nm beam. This SHG stage (SHG-1 in Fig. 1) is deliberately configured to have a relatively low conversion efficiency to take advantage of how the SHG process reshapes the transmitted fundamental beam (explained below).

Due to the relatively high gain and short pump duration of the NOPA stage, the pulse shaper cannot impose the full dispersion needed for the OPCPA chain. Thus, the near-IR NOPA output is further stretched with a prism pair providing additional negative GDD of -1400 fs$^2$ and TOD of -17000 fs$^3$ at 730 nm. Next, the mid-IR light is generated as an idler in a collinear 0.3-mm-long BBO-based difference frequency generation (DFG) stage. The output of the NOPA is focused into the DFG crystal to 80 GW/cm$^2$ peak intensity (estimated via the measured beam size and the calculated dispersion on the pulse). It is overlapped with part of the transmitted 1030-nm beam from SHG-1, focused to 12 GW/cm$^2$ peak intensity (estimated based on the measured beam size and the calculated SHG depletion from SHG-1). Because the DFG occurs between the shaped NOPA output and a fixed-phase narrowband seed, the phase defined by the pulse shaper is linearly transferred to the idler.

The generated mid-IR pulses are amplified in three periodically poled lithium niobate (PPLN) NOPA stages with type-0 phase-matching. For all of the amplification stages, we use PPLN samples with a high-aspect-ratio 2x10.9 mm$^2$ poled aperture and 29.3 µm poling period (HC Photonics Corp.). The length of the samples at each subsequent stage is 1 mm, 1 mm and 0.5 mm, respectively. Obtaining high gain in such short PPLN samples is enabled by their large nonlinear coefficient. This leads to an octave-wide phase-matching bandwidth, minimal mid-IR light absorption in the crystals, and mitigates photorefractive effects due to the high-power pump pulses. Furthermore, the PPLN crystals are wrapped in indium foil and placed in water-cooled copper mounts to ensure good thermal and electrical contact.

Each mid-IR frequency conversion stage leverages the spatiotemporal flattening effects of SHG on the transmitted fundamental wave (FW). This shaping occurs because the highest-intensity parts of the input pulse are converted to the second harmonic (SH) most efficiently, thereby providing a flattening effect in both space and time on the FW. Such flattening is advantageous for mitigating back-conversion effects in subsequent frequency conversion stages.

The first PPLN NOPA stage is pumped by the transmitted FW beam from SHG-1. The second and third stages use the same principle but have a different, separately adjustable SHG crystal (SHG-2, Fig. 1). While the SH output of SHG-1 pumps the BBO NOPA, the SH output of SHG-2 is simply discarded, as illustrated in Fig. 1. To preserve the flattened beam shape and also to ensure stability of the system, the pump beams from the SHG stages are imaged with cylindrical telescopes onto the PPLN crystals at estimated peak intensities of 23, 18 and 25 GW/cm$^2$, respectively. The beam flattening plays an important role in the operation of the system since it allows to increase the useful pump energy on the crystals and helps preventing early saturation of the amplification. Such operation conditions of the PPLN-based amplification stages permit using 100-W class pump powers with manageable thermal and photo-refractive distortions on the amplified signal beams.

The last amplification stage delivers up to 30 W of output signal power with a power slope plotted in Fig. 2a. The output beam is spatially shaped with cylindrical CaF$_2$ anti-reflection coated telescopes, spectrally filtered from any parasitic co-propagating light by multiple reflections on high-reflectivity ultra-broadband mirrors based on Si/SiO$_2$ multilayers (Optoman), and is finally compressed in a 25-mm-thick zinc selenide (ZnSe) bulk compressor oriented at Brewster's angle. This leads to 25 W compressed mid-IR output power available for experiments. ZnSe was chosen to compress the pulses due to its favorable GDD-to-TOD ratio. Minimum overall TOD in the system helps to maintain a monotonic chirp of the pulses, which prevents cross-talk between the spectral components in the amplification stages.

The compressed pulses were characterized using a third harmonic frequency-resolved optical gating technique (FROG). Figures 2b and 2c show the retrieved pulse duration and spectrum compared with an independently measured spectrum. The retrieved spectral phase is virtually flat over the full spectral range. The retrieved pulse shape yields a transform-limited 16.5-fs full-width at half-maximum (FWHM) duration. The peak power of the pulses was calculated to be 14.2 GW, which corresponds to the highest peak power reported for high-repetition-rate systems. The central wavelength of the output spectrum was calculated to be 2.2 µm and thus the pulse envelope contains only 2.2 oscillation cycles of the carrier wave at FWHM.

Compared to our previous mid-IR OPCPA result [20], this system operates in a regime of opposite GDD sign, which allowed to decrease the overall TOD. The system also contains one amplification stage less while providing a higher conversion efficiency. In particular, the quantum conversion efficiency in the final stage is as high as 41%. The spatiotemporal flattening of the pump beam for the PPLN-based NOPA stages allowed to maximize energy extraction from the pump while keeping a broadband amplification thanks to the improved pump pulse temporal shape.

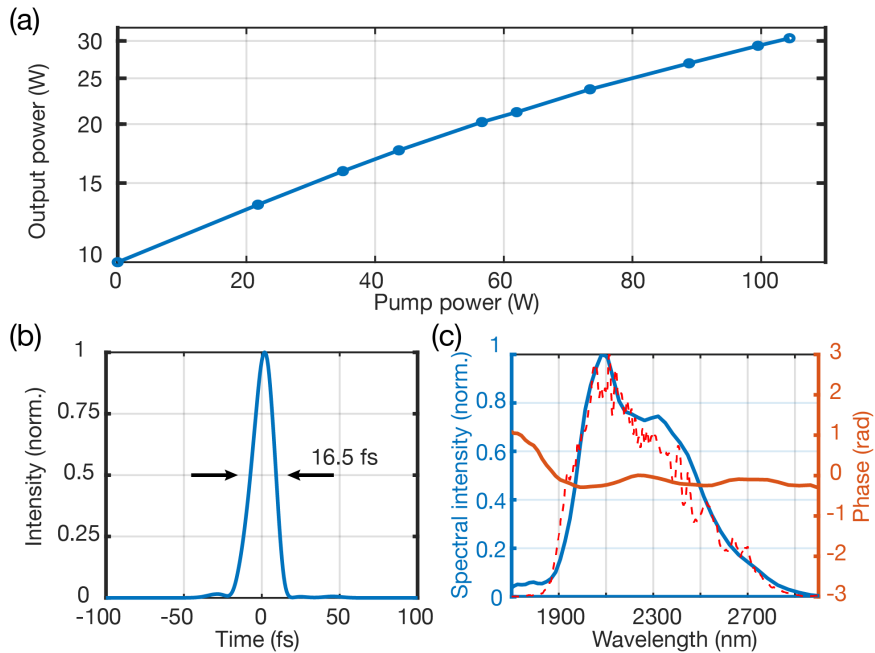

Figure 2. (a) The power slope of the last amplification stage. (b) Retrieved pulse duration of the amplifier output. (c) Blue line – retrieved spectrum, red-dashed line – measured spectrum, orange line – retrieved phase of the output pulses.

For high-harmonic generation experiments, the output beam of our mid-IR OPCPA is routed via a periscope system to another laboratory more than 15 meters away from the output of the system. The beam is guided by high-reflectivity Si/SiO$_2$ mirrors (Optoman) and its pointing is stabilized on another optical table using a stabilization unit (TEM Messtechnik GmbH). The full beam path including the OPCPA system itself is immersed in nitrogen gas to minimize water absorption.

For ionizing helium, the beam was tightly focused with a 75 mm long focal length plano-convex CaF$_2$ lens. The tight focusing leads to a strong Gouy-phase sweep through the interaction length. On top of that, the refractive index of helium at mid-IR and SXR wavelengths is close to unity. Thus, a high gas density is necessary to compensate the Gouy-phase and ionization of helium in order to achieve phase-matching conditions between the mid-IR and the SXR radiation [15]. From our calculations, we have found that a target pressure of 20-50 bar is necessary to phase-match mid-IR driven HHG in helium. A capillary-based approach, as demonstrated in [5,15], was found not to be a practical route in our case. The limited pulse energy available at this high repetition rate would require a small capillary, leading to high guiding losses. For this reason, we chose to have our gas target in a steel needle transversally drilled by the laser itself.

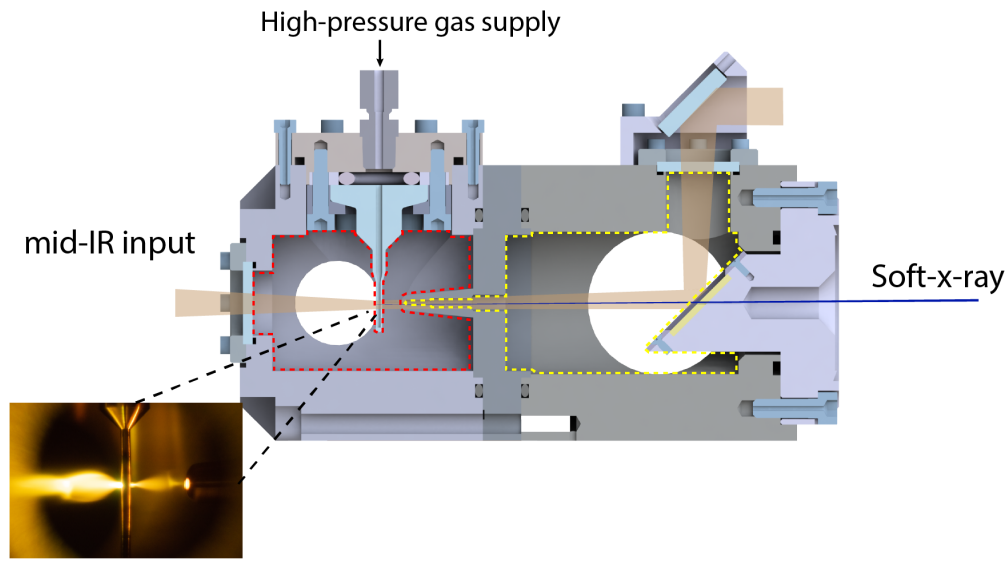

Figure 3. High-harmonic generation (HHG) with differential gas pumping. Red dashed path marks the first differential pumping stage and yellow dashed path marks the second one. The inset shows a photograph taken during the HHG process.

To ensure that high vacuum can be maintained after the HHG despite the required gas pressure, a differential pumping scheme was implemented (see Fig. 3). Our HHG module uses two differential pumping stages. The first stage (red dashed path) is pumped by a high-gas-load pump (A100L, Pfeiffer) while the second stage (yellow dashed path) is already supported by a turbo-molecular pump (HP300, Pfeiffer). The pressure in the stages was defined by the hole size in the needle. For the 1-mm-long target with two mechanically drilled 60-µm-diameter holes, we could supply more than 70 bar target gas pressure. At such conditions, the pressure is 7 mbar in the first, and $1.5 \cdot 10^{-2}$ mbar in the second stage, respectively.

The HHG module output is directly connected to a larger vacuum chamber with a residual gas pressure below $10^{-6}$ mbar for HHG characterization with a CCD-based flat-field spectrometer (251 MX, McPherson). The spectra were measured after blocking the residual pump light with a 100 nm thick aluminium filter. The spectrometer was calibrated on the observed absorption edges and by the position of harmonics where they are resolved. The C-edge is visible in all measurements and originates from organic contaminants in the spectrometer. The measured spectra were corrected for the CCD response, grating reflectivity and the filter transmission.

Figure 4 shows our results with helium, demonstrating HHG spanning the water window spectral region. At 400 eV and 40 bar target pressure, the absorption length in helium gas is 1.5 mm. Thus, reabsorption in the gas does not significantly reduce the SXR flux. Furthermore, to achieve absorption-limited HHG [26], the phase-matched interaction length should extend up to 8 mm. However, from spatiotemporal phase-matching calculations based on PPT ionization rates [27], we find that the maximum attainable coherence length of the HHG is only 1 mm for our parameters.

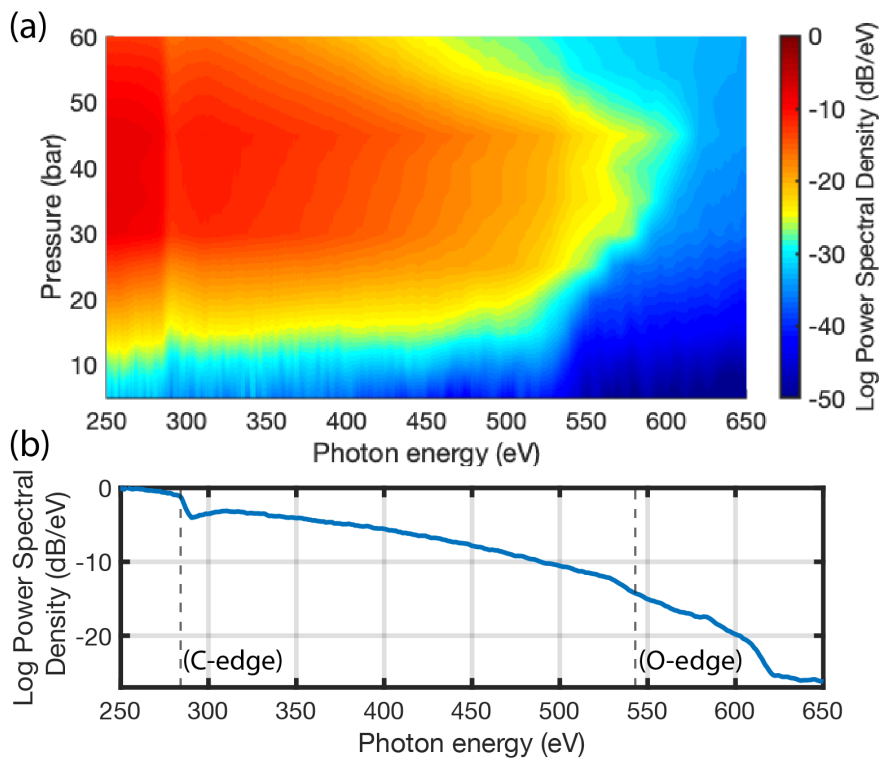

Figure 4. (a) The measured high-harmonic generation (HHG) spectrum from helium target gas as a function of the target pressure. (b) The measured HHG spectrum at 45 bar helium gas pressure.

For experiments where lower photon energies are desired and higher photon flux is necessary, HHG can be performed in gases with lower ionization potential. Figure 5 shows our experimental results for HHG from (a) argon and (b) neon. Compared to helium, neon HHG results in an order of magnitude higher flux, but a slightly lower cut-off of 415 eV. The achieved cut-off from argon gas was 200 eV with 172 times higher photon flux than from helium. The absolute flux of the HHG was not measured due to the lack of a suitable calibrated measurement instruments, but from the CCD counts the lower bound of the flux could be estimated to be 5.5 nW (60-200 eV) for argon, 300 pW for neon (220-420 eV) and 32 pW (250-620 eV) for helium spectra without considering energy loss on the spectrometer entrance slit.

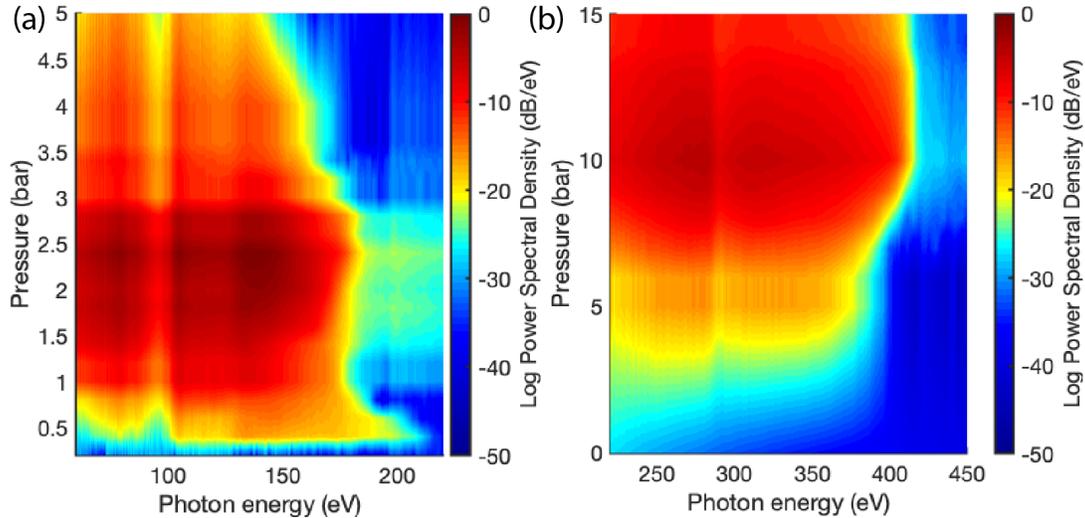

Figure 5. The measured high-harmonic generation (HHG) spectrum from (a) argon and (b) neon gas as a function of pressure.

In conclusion, we have demonstrated a 100-kHz mid-IR laser source delivering pulses with peak power over 14 GW and average power of 25 W. To the best of our knowledge, this corresponds to the world record output parameters for high-repetition rate mid-IR systems. The two-cycle pulses from the mid-IR system were used to generate a SXR continuum spanning the complete water window up to 0.6 keV. This proof-of-principle demonstration paves the way for high-repetition-rate experiments in the water window range. This laser source will enable a new generation of attosecond studies, where high repetition rates and high photon energies are combined to explore new regimes with space-charge-free ultrafast photoemission and coincidence detection techniques.

**Funding**
Swiss National Science Foundation (SNSF) (200020_172644), SNSF R'Equip (206021_164034/1).

**Acknowledgments**
We thank Marcel Baer and Andreas Stuker with their teams for support in preparing custom components and vacuum chambers. We also thank Benjamin Willenberg, Fabian Brunner and Stefan Hrisafov for their advice and support in the lab.